\documentclass{article}
\usepackage{spconf,amsmath,amssymb,graphicx}
\usepackage{color}
\usepackage{cite}
\usepackage{balance}

\usepackage[]{algorithm2e}
\usepackage{wasysym}
\usepackage{bm}
\usepackage{balance}
\usepackage{enumitem} %
\usepackage{pifont}%
\usepackage{xspace}
\usepackage{url}

\def\RR{\mathbb R}

\def\bsse{\texttt{BSS{\textunderscore}eval}\xspace}
\def\bei{\texttt{bss{\textunderscore}eval{\textunderscore}images}\xspace}
\def\bes{\texttt{bss{\textunderscore}eval{\textunderscore}sources}\xspace}
\def\bdg{\texttt{bss{\textunderscore}decomp{\textunderscore}gain}\xspace}

\newfont{\bg}{cmr10 scaled\magstep4}

\newcommand{\bigzerou}{\smash{\lower1.7ex\hbox{\bg 0}}}

\DeclareMathOperator*{\argmin}{argmin}

\title{SDR -- half-baked or well done?}
\name{Jonathan Le Roux$^1$, Scott Wisdom$^2$, 
Hakan Erdogan$^3$, John R. Hershey$^2$}
\address{$^1$Mitsubishi Electric Research Laboratories (MERL), Cambridge, MA, USA\\
$^2$Google AI Perception, Cambridge, MA  \hspace{1cm}
$^3$Microsoft Research, Redmond, WA} %

\begin{document}
\ninept
\maketitle
\setlength{\abovedisplayskip}{4pt}
\setlength{\belowdisplayskip}{4pt}
\begin{abstract}
In speech enhancement and source separation, signal-to-noise ratio is a ubiquitous objective measure of denoising/separation quality. A decade ago, the \bsse toolkit was developed to give researchers worldwide a way to evaluate the quality of their algorithms in a simple, fair, and hopefully insightful way: it attempted to account for channel variations, and to not only evaluate the total distortion in the estimated signal but also split it in terms of various factors such as remaining interference, newly added artifacts, and channel errors. In recent years, hundreds of papers have been relying on this toolkit to evaluate their proposed methods and compare them to previous works, often arguing that differences on the order of 0.1 dB proved the effectiveness of a method over others. We argue here that the signal-to-distortion ratio (SDR) implemented in the \bsse toolkit has generally been improperly used and abused, especially in the case of single-channel separation, resulting in misleading results. We propose to use a slightly modified definition, resulting in a simpler, more robust measure, called scale-invariant SDR (SI-SDR). We present various examples of critical failure of the original SDR that SI-SDR overcomes.
\end{abstract}

\section{Introduction}
\label{sec:intro}
Source separation and speech enhancement have been an intense focus of research in the signal processing community for several decades, and interest has gotten even stronger with the recent advent of powerful new techniques based on deep learning \cite{lu2013speech,Weninger2014RNN,xu2014experimental,Erdogan2015ICASSP04,weninger2015speech, Hershey2016ICASSP03,Isik2016Interspeech09,Yu2017PIT,kolbaek2017uPIT,Wang2017Overview,Wang2018ICASSP04Alternative}. An important area of research has focused on single-channel methods, which can denoise speech or separate one or more sources from a mixture recorded using a single microphone.
Many new methods are proposed, and their relevance is generally justified by their outperforming some previous method according to some objective measure.

While the merits of various objective measures such as PESQ \cite{rix2001perceptual}, Loizou's composite measure \cite{Loizou2007}, PEMO-Q \cite{huber2006pemo}, PEASS \cite{Emiya2011subjective}, or STOI \cite{taal2010short}, could be debated and compared, we are concerned here with an issue with the way the widely relied upon \bsse toolbox \cite{Vincent2006BSSeval} has been used. We focus here on the single-channel setting. The \bsse toolbox reports objective measures related to the signal-to-noise ratio (SNR), attempting to account for channel variations, and to report a decomposition of the overall error, referred to as signal-to-distortion ratio (SDR), into components indicating the type of error:  source image to spatial distortion ratio (ISR), signal to interference ratio (SIR), and signal to artifacts ratio (SAR).
In version 3.0, \bsse featured two main functions, \bei and \bes. 
\begin{itemize}[leftmargin=*]
\item \bes completely forgives channel errors that can be accounted for by a time-invariant 512-tap filter, modifying the reference to best fit each estimate. This includes very strong modifications of the signal, including low-pass or high-pass filters. Thus, obliterating some frequencies of a signal by setting them to 0 could absurdly still result in near infinite SDR.
\item \bei reports channel errors (including gain errors) as errors in the ISR measure, but its SDR is nothing else than vanilla SNR. While not as fatal as the modification of the reference in \bes, \bei suffers from some issues. First, it does not even allow for a global rescaling factor, which may occur when one tries to avoid clipping in the reconstructed signal. Second, as does SNR, it takes the scaling of the estimate at face value, a loophole that algorithms could (potentially unwittingly) exploit, as explained in section \ref{sec:SNR}.
\end{itemize}

An earlier version (2.1) of the toolbox does provide, among other functions, a decomposition which only allows a constant gain via the function \bdg. Performance criteria such as SDR can then be computed from this decomposition, but most papers on single-channel separation appear to be using \bes. 
The \bsse website\footnote{\url{http://bass-db.gforge.inria.fr/bss_eval/}} actually displays a warning about which version should be used. Version 3.0 ``{\it is recommended for mixtures of reverberated or diffuse sources (aka convolutive mixtures), due to longer decomposition filters enabling better correlation with subjective ratings. It [is] also recommended for instantaneous mixtures when the results are to be compared with SiSEC.}'' On the other hand, version 2.1 ``{\it is practically restricted to instantaneous mixtures of point sources. It is recommended for such mixtures, except when the results are to be compared with SiSEC.}'' It appears that this warning has not been understood, and most papers use Version 3.0 without further consideration. The desire to compare results to (early editions of) SiSEC should also not be a justification for using a flawed measure.
The same issues apply to an early Python version of \bsse{}, \texttt{bss{\textunderscore}eval}\footnote{\url{http://github.com/craffel/mir_eval/}} \cite{raffel2014mir_eval}. Recently, \bsse v4 was released as a Python implementation\footnote{\url{https://sigsep.github.io/sigsep-mus-eval/museval.metrics.html}} \cite{stoter20182018}: the authors of Version 4 acknowledged the issue with the original \bes, and recommended using \bei instead. This however does not address the scaling issue.

These problems shed doubt on many results, including some in our own older papers, especially in cases where algorithms differ by a few tenths of a dB in SDR.
This paper is intended both to illustrate and propagate this message more broadly, and also to encourage the use, for single-channel separation evaluation, of simpler, scale-aware, versions of SDR: scale-invariant SDR (SI-SDR) and scale-dependent SDR (SD-SDR). 
We also propose a definition of SIR and SAR  in which there is a direct relationship between SDR, SIR, and SAR, which we believe is more intuitive than that in \bsse. 
The scale-invariant SDR (SI-SDR) measure was used in \cite{Hershey2016ICASSP03,Isik2016Interspeech09,Luo2017,Wang2018ICASSP04Alternative,Wang2018Interspeech09,Chen2017,Luo2018TasNet09arXiv}.  Comparisons in \cite{Wang2018Interspeech09} showed that there is a significant difference between SI-SDR  and the SDR as implemented in \bsse's \bes function.  
We review the proposed measures, show some critical failure cases of SDR, and give a numerical comparison on a speech separation task.

\section{Proposed measures}
\subsection{The problem with changing the reference}
A critical assumption in \bes, as it is implemented in the publicly released toolkit up to Version 3.0, is that time-invariant filters are considered allowed deformations of the target/reference. One potential justification for this is that a reference may be available for a source signal instead of the spatial image at the microphone which recorded the noisy mixture, and that spatial image is likely to be close to the result of the convolution of the source signal with a short FIR filter, as an approximation to its convolution with the actual room impulse response (RIR). This however leads to a major problem, because the space of signals achievable by convolving the source signal with any short FIR filter is extremely large and includes perceptually widely different signals from the spatial image.
Note that the original \bsse paper \cite{Vincent2006BSSeval} also considered time-varying gains and time-varying filters as allowed deformations. Taken to an extreme, this creates the situation where the target can be deformed to match pretty much any estimate.

Modifying the target/reference when comparing algorithms is deeply problematic when the modification depends on the outputs of each algorithm.
In effect, \bes chooses a different frequency weighting of the error function depending on the spectrum of the estimated signal: frequencies that match the reference are emphasized, and those that do not are discarded.  Since this weighting is different for each algorithm, \bes cannot provide a fair comparison between algorithms.

\vspace{-.2cm}
\subsection{The problem with not changing anything}
\label{sec:SNR}

\begin{figure}[t]
	\centering
	\includegraphics[width=.8\columnwidth]{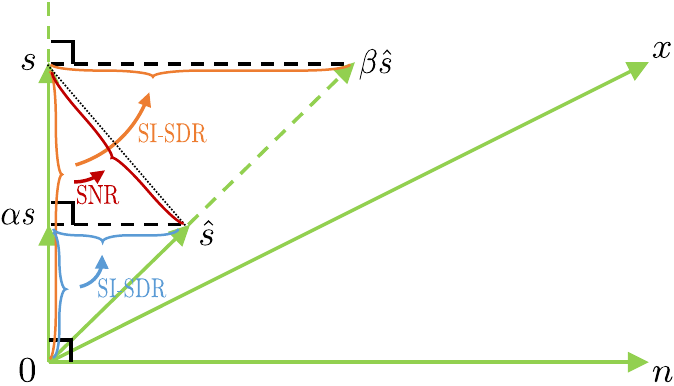}\vspace{-.2cm}
	\caption{Illustration of the definitions of SNR and SI-SDR.}\vspace{-.4cm}
	\label{fig:projection}
\end{figure}

Let us consider a mixture $x=s+n \in \RR^L$ of a target signal $s$ and an interference signal $n$. 
Let $\hat{s}$ denote an estimate of the target obtained by some algorithm. The classical SNR (which is equal to \bei's SDR) considers $\hat{s}$ as the estimate and $s$ as the target:
\begin{align}
	\text{SNR} &= 10 \log_{10} \left( \frac{||s||^2}{||s - \hat{s}||^2}  \right).
\end{align}
As is illustrated in Fig.~\ref{fig:projection}, where for simplicity we consider the case where the estimate is in the subspace spanned by speech and noise (i.e., no artifact), what is considered as the noise in such a context is the residual $s-\hat{s}$, which is not guaranteed to be orthogonal to the target $s$. A tempting mistake is to artificially boost the SNR value without changing anything perceptually by rescaling the estimate, for example to the orthogonal projection of $s$ on the line spanned by $\hat{s}$: this leads to a right triangle whose hypotenuse is $s$, so SNR could always be made positive. In particular, starting from a mixture $x$ where $s$ and $n$ are orthogonal signals with equal power, so with an SNR of 0 dB, projecting $s$ orthogonally onto the line spanned by $x$ corresponds to rescaling the mixture to $x/2$: this ``improves'' SNR by $3$ dB. Interestingly, \bei's ISR is sensitive to the rescaling, so the ISR of $x$ will be higher than that of $x/2$, while its SDR (equal to SNR for \bei) is lower.

\subsection{Scale-aware SDR}

To ensure that the residual is indeed orthogonal to the target, we can either rescale the target or rescale the estimate. Rescaling the target such that the residual is orthogonal to it corresponds to finding the orthogonal projection of the estimate $\hat{s}$ on the line spanned by the target $s$, or equivalently finding the closest point to $\hat{s}$ along that line.
This leads to two equivalent definitions for what we call the scale-invariant signal-to-distortion ratio (SDR):
\begin{align}
\text{SI-SDR}&=\frac{|s|^2}{| s - \beta \hat{s}|^2} \text{ for }\beta \text{ s.t.\ } s \perp s-\beta\hat{s} \\
&=\frac{|\alpha s|^2}{|\alpha s - \hat{s}|^2}\text{ for }\alpha = \argmin_\alpha{|\alpha s - \hat{s}|^2}.
\end{align}
The optimal scaling factor for the target is obtained as $\alpha=\hat{s}^T s / ||s||^2$, and the scaled reference is defined as $e_\text{target}= \alpha s$. We then decompose the estimate $\hat{s}$ as $\hat{s} = e_\text{target} + e_\text{res}$, leading to the expanded formula:
\begin{align}
\text{SI-SDR} &= 10 \log_{10} \left( \frac{||e_\text{target}||^2}{||e_\text{res}||^2}  \right)\\
&=10 \log_{10} \left( \frac{||\frac{\hat{s}^T s}{||s||^2} s||^2}{ ||\frac{\hat{s}^T s}{||s||^2} s - \hat{s}||^2 }  \right).
\end{align}
Instead of a full 512-tap FIR filter as in \bsse, SI-SDR uses a single coefficient to account for scaling discrepancies. As an extra advantage, computation of SI-SDR is thus straightforward and much faster than that of SDR. Note that SI-SDR corresponds to the SDR obtained from \bdg in \bsse Version 2.1. %
SI-SDR has recently been used as an objective measure in the time domain to train deep learning models for source separation, outperforming least-squares on some tasks \cite{Venkataramani2018arxiv,Luo2018TasNet09arXiv} (it is referred to as SDR in \cite{Venkataramani2018arxiv} and as SI-SNR in \cite{Luo2018TasNet09arXiv}).

A potential drawback of SI-SDR is that it does not consider scaling as an error. In situations where this is not desirable, one may be interested in designing a measure that does penalize rescaling. Doing so turns out not to be straightforward. As we saw in the example in Section~\ref{sec:SNR} of a mixture $x$ of two orthogonal signals $s$ and $n$ with equal power, considering the rescaled mixture $\hat{s}=\mu x$ as the estimate, SNR does not peak at $\mu=1$ but instead encourages a down-scaling of $\mu=1/2$. It does however properly discourage large up-scaling factors. As an alternative measure that properly discourages down-scalings, we propose a scale-dependent SDR (SD-SDR), where we consider the rescaled $s$ as the target $e_\text{target}=\alpha s$, but consider the total error as the sum of two terms,  $||\alpha s - \hat{s}||^2$ accounting for the residual energy, and $|| s - \alpha s||^2$ accounting for the rescaling error. Because of orthogonality, $||\alpha s - \hat{s}||^2 + || s - \alpha s||^2 = ||s - \hat{s}||^2$, and we obtain:
\begin{align}
\text{SD-SDR} &= 10 \log_{10} \left( \frac{||\alpha s||^2}{||s - \hat{s}||^2}  \right)=\text{SNR} + 10 \log_{10} \alpha^2
\end{align}
Going back to the example in Section~\ref{sec:SNR}, SI-SDR is independent of the rescaling of $x$, while SD-SDR for $\hat{s}=\mu x$ is equal to 
\begin{align}
10 \log_{10} \left( \frac{\|\mu s\|^2}{\| s - \mu x \|^2}\right) &= 10 \log_{10} \left( \frac{\mu^2 \|s\|^2}{\| (1-\mu) s - \mu n \|^2}\right) \\
&= 10 \log_{10} \left( \frac{\mu^2}{ (1-\mu)^2 + \mu^2}\right),
\end{align}
which does peak at $\mu=1$. While this measure properly accounts for down-scaling errors where $\mu<1$, it only decreases to $-3$ dB for large up-scaling factors $\mu \gg 1$. 
For those applications where both down-scaling and up-scaling are critical, one could consider the minimum of SNR and SD-SDR as a relevant measure.

\vspace{-.2cm}
\subsection{SI-SIR and SI-SAR}

In the original \bsse toolkit, the split of SDR into SIR and SAR is done in a mathematically non intuitive way: in the original paper, the SAR is defined as the ``sources to artifacts ratio,'' not the ``source to artifacts ratio,'' where ``sources'' refers to all sources, including the noise. That is, if the estimate contains more noise, yet everything else stays the same, then the SAR actually goes up. There is also no simple relationship between SDR, SIR, and SAR.

Similarly to \bsse, we can further decompose  $e_\text{res}$ as
$e_\text{res} = e_\text{interf} + e_\text{artif}$, where $e_\text{interf}$ is defined as the orthogonal projection of  $e_\text{res}$ onto the subspace spanned by both $s$ and $n$. 
But differently from \bsse, we define the scale-invariant signal  to  interference  ratio  (SI-SIR)  and  the scale-invariant  signal  to  artifacts ratio (SI-SAR) as follows:
\begin{align}
\text{SI-SIR} &= 10 \log_{10} \left( \frac{||e_\text{target}||^2}{||e_\text{interf}||^2}  \right),\\
\text{SI-SAR} &=10 \log_{10} \left( \frac{||e_\text{target}||^2}{||e_\text{artif}||^2}  \right).
\end{align}
These definitions have the advantage over those of \bsse that they verify
\begin{equation}
    10^{-\text{SI-SDR}/10}=10^{-\text{SI-SIR}/10}+10^{-\text{SI-SAR}/10},
\end{equation} 
because the orthogonal decomposition leads to $||e_\text{res}||^2 = ||e_\text{interf}||^2 + ||e_\text{artif}||^2 $. There is thus a direct relationship between the three measures. Scale-dependent versions can be defined similarly.

That being said, we feel compelled to note that, whether it is still relevant to split SDR into SIR and SAR is a matter of debate: machine-learning based methods tend to perform a highly non-stationary type of processing, and using a global projection on the whole signal may thus not be guaranteed to provide the proper insight.

\vspace{-.2cm}
\section{Examples of extreme failure cases}

We present some failure modes of SDR that SI-SDR overcomes.
\vspace{-.2cm}
\subsection{Optimizing a filter to minimize SI-SDR}
\label{sec:optimize_filter}

For this example, we optimize an STFT-domain, time-invariant filter to minimize SI-SDR. We will show that despite SI-SDR being minimized by the filter, SDR performance remains relatively high since it is allowed to apply filtering to the reference signal.

Optimization of the filter that minimizes SI-SDR is implemented in Keras with a Tensorflow backend, where the trainable weights are an $F$-dimensional vector ${\bf w}$. A sigmoid nonlinearity is applied to this vector to ensure the filter has values between 0 and 1, and the final filter ${\bf m}$ is obtained by renormalizing ${\bf v}=\mathrm{sigm}({\bf w})$ to have unit $\ell_\infty$-norm:
${\bf m}={\bf v} / \|{\bf v}\|_\infty$. The filter is optimized on a single speech example using gradient descent, where the loss function being minimized is SI-SDR. Application of the masking filter is implemented end-to-end, where gradients are backpropagated through an inverse STFT layer.

\begin{figure}[t]
	\centering
	\includegraphics[width=0.8\columnwidth]{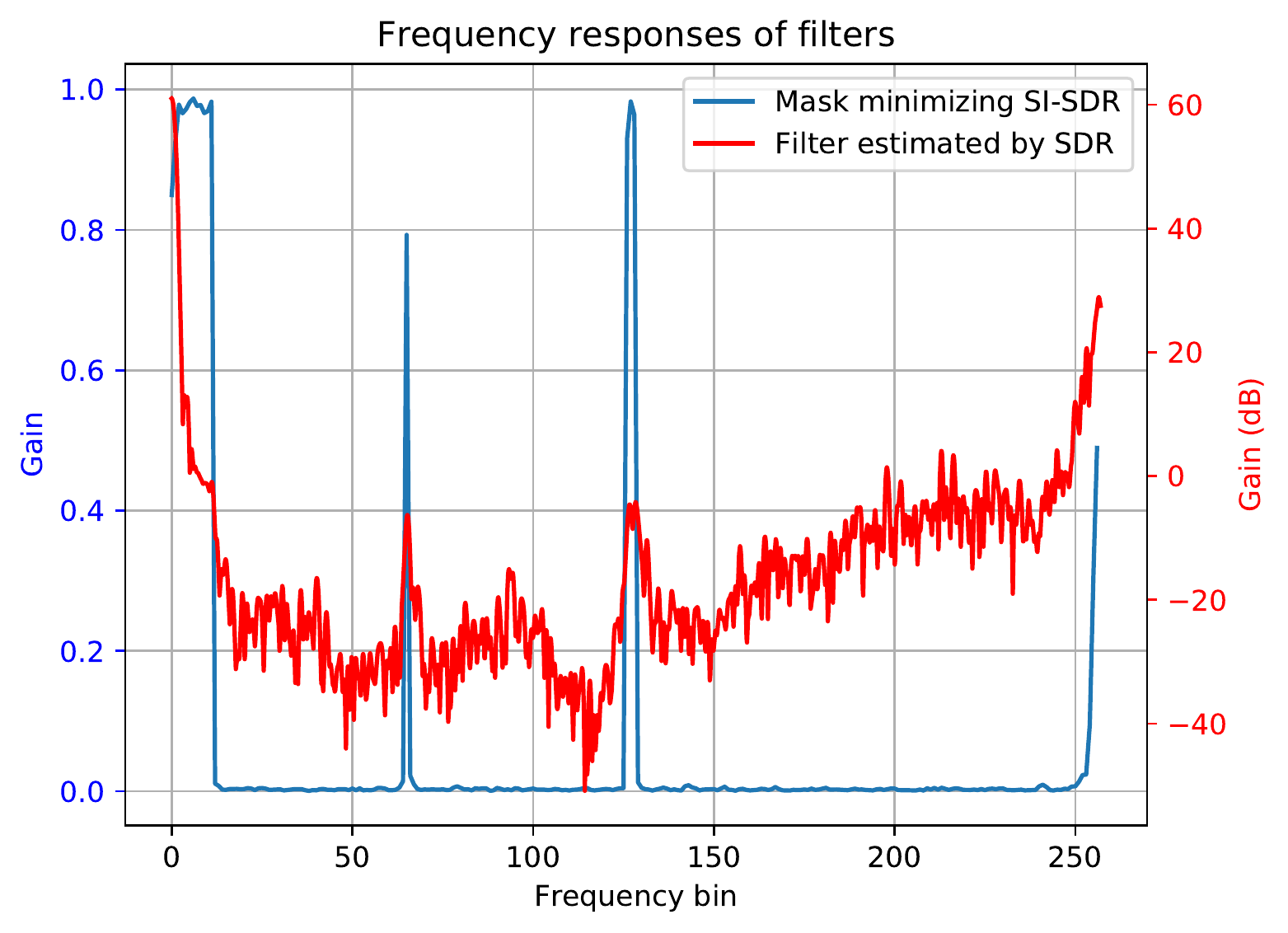}
	\vspace{.2cm}
	\includegraphics[width=0.94\columnwidth]{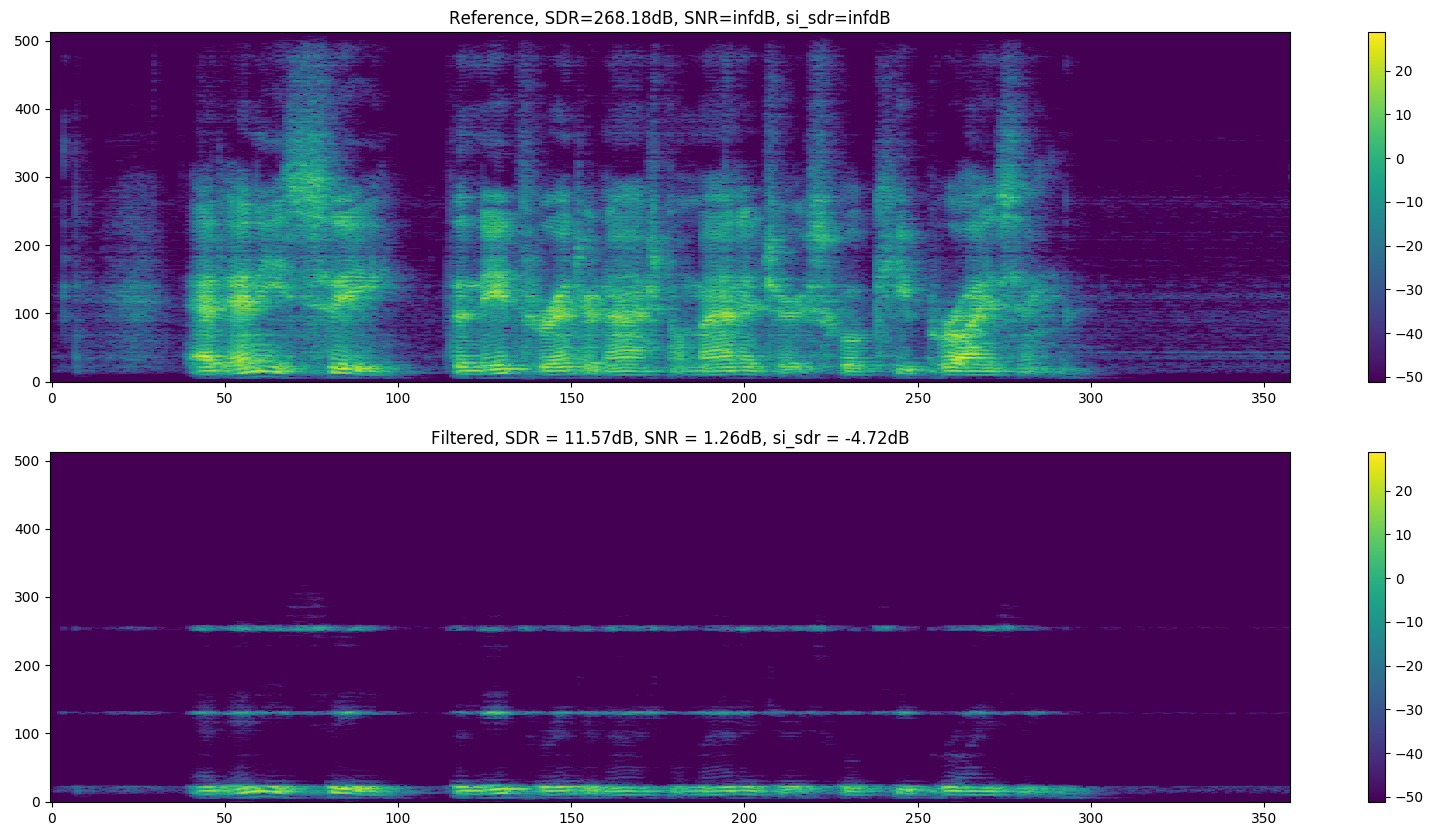}
	\caption{Top: filter applied to a clean speech signal that minimizes SI-SDR (blue) and magnitude response of the FIR filter estimated by SDR (red). Bottom: spectrograms of a clean speech signal (top) and the same signal processed by the optimized filter in blue above. %
	}	\vspace{-.2cm}
	\label{fig:filter}
\end{figure}

An example of a learned filter and resulting spectrograms for a single male utterance from CHiME2 is shown in Fig.~\ref{fig:filter}. To minimize SI-SDR, the filter learns to remove most of the signal's spectrum, only passing a couple of narrow bands. This filter achieves -4.7 dB SI-SDR, removing much of the speech content. However, despite this destructive filtering, we have the paradoxical result that the SDR of this signal is still high at 11.6 dB, since \bsse is able to find a filter to be applied to the reference signal that removes similar frequency regions. This filter is shown in red in the top part of Fig.~\ref{fig:filter}, somewhat matching the filter minimizing SI-SDR in blue.   %

\vspace{-.2cm}
\subsection{Progressive deletion of frequency bins}

The previous example illustrated that SDR can yield high scores despite large regions of a signal's spectrum being deleted. Now we examine how various metrics perform when frequency bins are progressively deleted from a signal.

\begin{figure}
    \centering
    \includegraphics[width=0.8\linewidth]{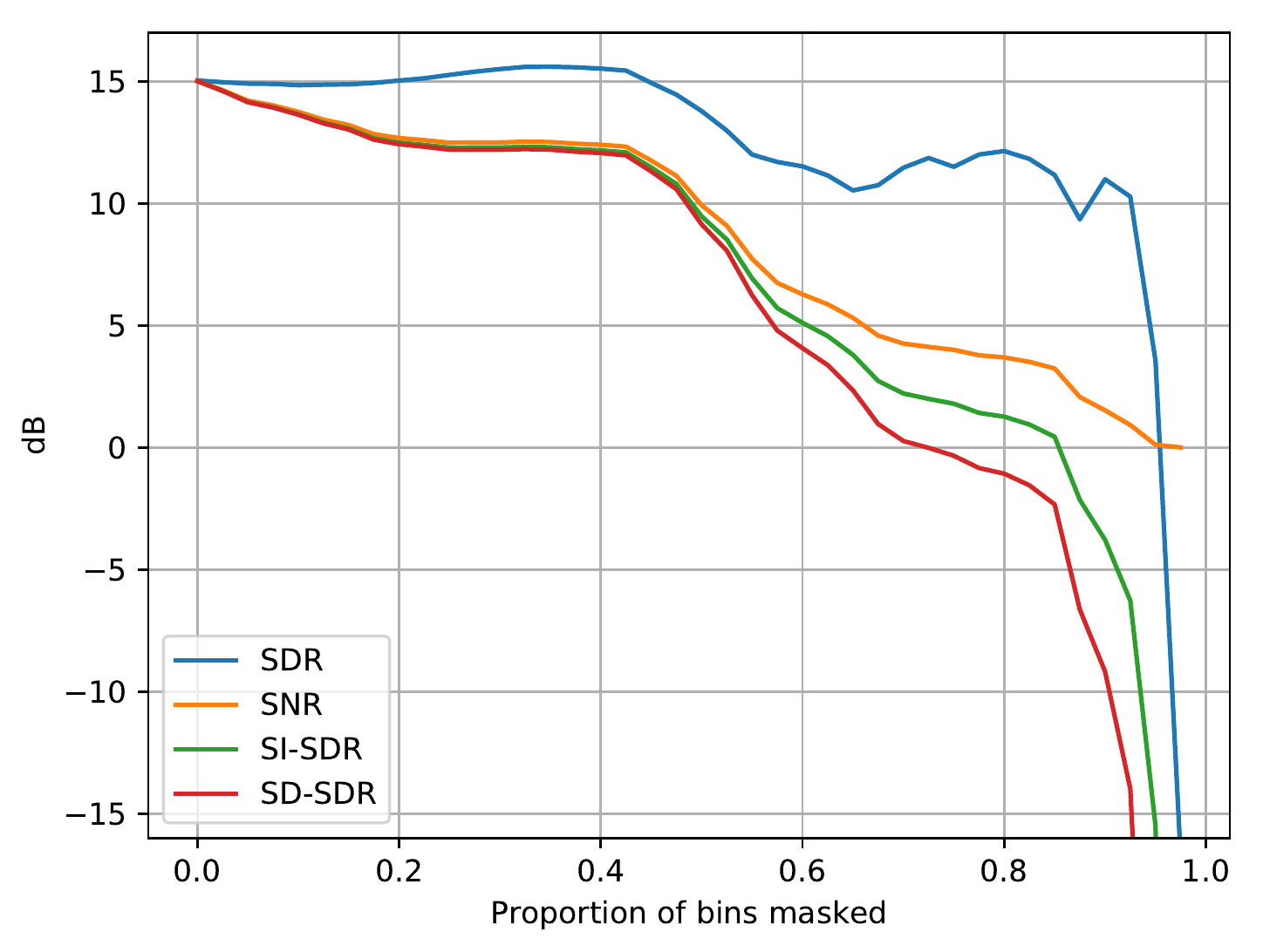}
    \vspace{-.2cm}
    \caption{Various metrics plotted versus proportion of frequency bins deleted for a speech signal plus white noise at 15dB SNR. %
    }\vspace{-.3cm}
    \label{fig:progdel}
\end{figure}

We add white noise at 15 dB SNR to the same speech signal used in Section~\ref{sec:optimize_filter}. Then time-invariant STFT-domain masking is used to remove varying proportions of frequency bins, where the mask is bandpass with a center frequency at the location of median spectral energy of the speech signal averaged across STFT frames. We measure four metrics: SDR, SNR, SI-SDR, and SD-SDR. The results are shown in Fig.~\ref{fig:progdel}. Despite more and more frequency bins being deleted, SDR  (blue) remains between 10 dB and 15 dB, until nearly all frequencies are removed. In fact, SDR even {\it increases} for a masking proportion of 0.4. In contrast, the other metrics more appropriately measure signal degradation since they monotonically decrease.

An important practical scenario in which such behavior would be fatal is that of bandwidth extension: it is not possible to properly assess the baseline performance, where upper frequency bins are silent, using SDR.

\subsection{Varying band-stop filter gain for speech corrupted with band-pass noise}

In this example, we consider adding bandpass noise to a speech signal, then applying a mask that filters the noisy signal in this band with varying gains, as a crude representation of a speech enhancement task. We mix the speech signal with a bandpass noise signal, where the local SNR within the band is 0 dB, and the band is 1600 Hz wide (20\% of the total bandwidth for a sampling frequency of 16 kHz), centered at the maximum average spectral magnitude across STFT frames of the speech signal. In this case, the optimal time-invariant Wiener filter should be bandstop, with a gain of 1 outside the band and a gain of about 0.5 within the band,
since the speech and noise have approximately equal power, and the Wiener filter is $P_\mathrm{speech} / (P_\mathrm{speech} + P_{noise}$).

\begin{figure}
    \centering
    \includegraphics[width=0.8\linewidth]{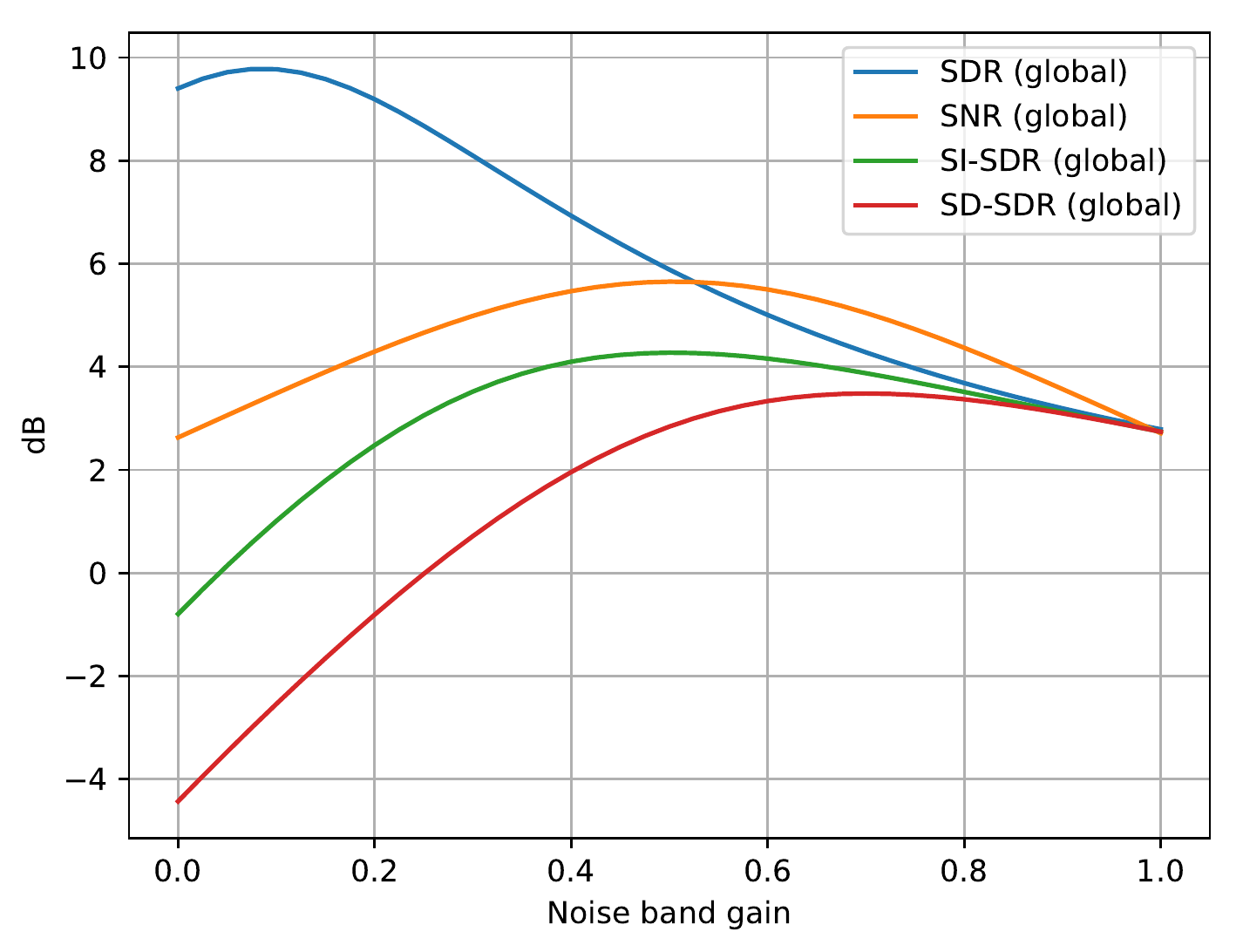}
    \vspace{-.28cm}
    \caption{Various metrics plotted versus bandstop filter gain for a speech signal plus bandpass white noise with 0dB SNR in the band. %
    }\vspace{-.3cm}
    \label{fig:varygain}
\end{figure}

We consider the performance of such filters when varying the bandstop gain from 0 to 1 in steps of 0.025, again for SDR, SNR, SI-SDR, and SD-SDR. The results are shown in Fig.~\ref{fig:varygain}. Notice that SNR, SI-SDR have a peak around a gain of 0.5 as expected. However, SDR monotonically increases as gain decreases. This is an undesirable behavior, as SDR becomes more and more optimistic about signal quality as more of the signal's spectrum is suppressed, because it is all too happy to see the noisy part of the spectrum being suppressed and modify the reference to focus only on the remaining regions. SD-SDR peaks slightly above 0.5, because it penalizes the down-scaling of the speech signal within the noisy band.

\section{Comparison on a speech separation task}

Both SI-SDR and \bsse's SDR have recently been used by various studies \cite{Hershey2016ICASSP03, Isik2016Interspeech09,Chen2017, Luo2018, Yu2017PIT, kolbaek2017uPIT,Wang2018ICASSP04Alternative,Wang2018Interspeech09,Luo2017a,Luo2018TasNet09arXiv} in the context of single-channel speaker-independent speech separation on the wsj0-2mix dataset \cite{Hershey2016ICASSP03}, some of these studies reporting both figures \cite{Chen2017, Luo2018,Wang2018Interspeech09,Luo2018TasNet09arXiv}. We gather in Table~\ref{table:comparison} various SI-SDR and \bsse SDR improvements (in dB) on the test set of the wsj0-2mix dataset mainly from \cite{Wang2018ICASSP04Alternative}, to which we add the recent state-of-the-art score of \cite{Luo2018TasNet09arXiv}.
The difference between the SI-SDR and the SDR scores for the algorithms considered are around 0.5 dB, but vary from 0.3 dB to 0.6 dB. Note furthermore that the algorithms considered here all result in signals that can be considered of good perceptual quality: much more varied results could be obtained with algorithms that give worse results. If the targets and interferences in the dataset were more stationary, such as in some speech enhancement scenarios, it is also likely there could be loopholes for SDR to exploit, where a drastic distortion that can be well approximated by a short FIR filter happens to lead to similar results on the mixture and the reference signals.

\begin{table}[t]
    \footnotesize
	\centering
	\caption{Comparison of improvements in SI-SDR and SDR for various speech separation systems on the wsj0-2mix dataset test set \cite{Hershey2016ICASSP03}.} \vspace{0.05cm}
	\label{table:comparison}
	\begin{tabular}{l|c|c}
		\hline\hline
		\multicolumn{1}{c|}{Approaches} & {SI-SDR [dB]} & {SDR [dB]} \\ 
		\hline\hline
		Deep Clustering \cite{Hershey2016ICASSP03, Isik2016Interspeech09} & 10.8 & - \\
		\hline
		Deep Attractor Networks \cite{Chen2017, Luo2018}  & 10.4 & 10.8 \\
		\hline
		PIT \cite{Yu2017PIT, kolbaek2017uPIT} & - & 10.0 \\
		\hline
        TasNet \cite{Luo2017a} & 10.2 & 10.5 \\
		\hline
        Chimera++ Networks \cite{Wang2018ICASSP04Alternative}  & 11.2  & 11.7 \\
        \quad + MISI-5 \cite{Wang2018ICASSP04Alternative}  & 11.5 & 12.0 \\
        \hline
        WA \cite{Wang2018Interspeech09}  & 11.8 & 12.3 \\
        WA-MISI-5 \cite{Wang2018Interspeech09}  & {12.6} & {13.1}\\
		\hline
		Conv-TasNet-gLN \cite{Luo2018TasNet09arXiv} & 14.6 & 15.0 \\
		\hline
		Oracle Masks: &&\\
		\quad Magnitude Ratio Mask  & 12.7  & 13.2 \\
		\quad \quad + MISI-5  & 13.7  & 14.3\\
		\quad Ideal Binary Mask  & 13.5  & 14.0 \\ 
		\quad \quad + MISI-5  & 13.4  & 13.8 \\
		\quad PSM  & 16.4  & 16.9 \\ 
		\quad \quad + MISI-5  & 18.3 & 18.8 \\
		\quad Ideal Amplitude Mask  & 12.8 & 13.2 \\ 
		\quad \quad + MISI-5  & 26.6 & 27.1 \\
		\hline\hline
	\end{tabular}\vspace{-0.3cm}
\end{table}

\section{Conclusion}
We discussed issues that pertain to the way \bsse's SDR measure has been used, in particular in single-channel scenarios, and presented a simpler scale-invariant alternative called SI-SDR. We also showed multiple failure cases for SDR that SI-SDR overcomes.

\vspace{0.2cm}
{\bf Acknowledgements:} 
The authors would like to thank Dr.\ Shinji Watanabe (JHU) and Dr.\ Antoine Liutkus and Dr.\ Fabian St{\"o}ter (Inria) for fruitful discussions.

\vfill\pagebreak
\balance

\bibliographystyle{IEEEtran_nourl}
\bibliography{SPL2017SDR,phasebook}

\end{document}